\documentclass[a4paper]{article}
\usepackage{mathrsfs}
\usepackage{graphicx}
\usepackage{latexsym}
\usepackage{amsmath}
\usepackage{amssymb}
\usepackage{textcomp}
\usepackage{amsbsy}
\usepackage{graphics}
\usepackage{color}
\usepackage{epstopdf}

\title{\large \bf Electromagnetic effect on anisotropic scalar field collapse in higher curvature gravity}
\author{Narayan Banerjee\footnote{E-mail address: narayan@iiserkol.ac.in}\\
Department of Physical Sciences,\\ 
Indian Institute of Science Education and Research Kolkata,\\
Mohanpur Campus, Nadia, West Bengal 741246, India.\\[10mm]
Tanmoy Paul\footnote{E-mail address: pul.tnmy9@gmail.com}\\
Department of Theoretical Physics,\\
Indian Association for the Cultivation of Science,\\
2A $\&$ 2B Raja S.C. Mullick Road,\\
Kolkata - 700 032, India.\\[10mm]}
\date{}
\begin{document}
\maketitle

\begin{abstract}
We consider a ``Scalar-Maxwell-Einstein-Gauss-Bonnet'' theory in four dimension, where the scalar field couples non-minimally with 
the Gauss-Bonnet (GB) term. This coupling with the scalar field ensures the non topological character 
of the GB term. In such higher curvature scenario, we explore the effect of electromagnetic field on scalar field collapse. 
Our results reveal that the presence of a time dependent electromagnetic field requires an anisotropy in the background 
spacetime geometry and such anisotropic spacetime allows a collapsing solution for the scalar field. 
The singularity formed as a result of the collapse is found to be a curvature singularity which may be point like or line 
like depending on the strength of the anisotropy. We also show that the singularity is always hidden from exterior by an apparent horizon.
\end{abstract}

\newpage

\tableofcontents

\section{Introduction}
Relativistic astrophysics have gone through extensive developments over a few decades, following the 
discovery of high energy phenomena in the universe such as gamma ray bursts. Interesting physical properties
can be emerged from compact stellar objects like neutron stars where the effect of strong gravity fields and 
hence general relativity are seen to play a 
fundamental role. The high strength of gravitational field is also present in the end stage of a continual 
gravitational collapse of a massive star. This collapsing phenomena, dominated by the force of gravity, 
is fundamental in black hole physics and have received increasing attention in the past decades. The 
first systematic analysis of gravitational collapse in general relativity was given way back in 1939 by 
Oppenheimer and Snyder \cite{oppenheimer} (see also the work by Datt \cite{datt}). Subsequent developments 
in the study of gravitational collapse have been comprehensively explored by Joshi\cite{joshi1,joshi2}.\\

Scalar fields have been of great interest in the gravity sector for its own reasons. The various forms of 
scalar field potential are good enough for cosmological requirements such as playing the role of the driver of the present 
or the early accelerated expansion of the universe. A suitable scalar potential can often mimic different equations of state 
of fluid distribution. Thus scalar fields, although hardly 
motivated by other branches of physics for its raison de etre, always enjoyed a lot of attention in gravitational physics.\\

Scalar fields are also quite popular in the context of collapsing spacetime geometry. The collapsing phenomena of a massless scalar field 
was discussed by Christodoulou \cite{1}. The possibilities of end product of a scalar field collapse, whether a naked singularity or a black hole, 
has also been explored in \cite{2}. The variants of scalar field collapse and its consequences 
are demonstrated in \cite{4,5,6,7,8,cai1,cai2,9,10,17,18} (see also \cite{soumya,nb4,13,14,15,16}).\\

The implementation of electromagnetic field in cosmological and astrophysical processes is an attractive research 
area in theoretical physics. Many investigations in this direction are devoted to understand the interaction between 
electromagnetic and gravitational fields. Bekenstein \cite{new1} extended the work from neutral to charged 
case by generalizing Oppenheimer-Volkoff equations \cite{new2} regarding the force balance of a star. Since then a considerable amount of work 
has been done in this scenario. Rosales et al. \cite{new3} figured out that electric charge plays the 
same role as that of anisotropy in the collapse, when the radial pressure is less than the 
tangential pressure. Thorne studied \cite{thorne} cylindrically symmetric gravitational collapse with 
magnetic field and concluded that magnetic field can prevent the collapse of cylinder before singularity formation. 
Ardavan and Partovi \cite{ardavan} investigated dust solution of the field equations with electromagnetic field and found that the 
electrostatic force is balanced by gravitational force during collapse of charged dust. Stein-Schabes \cite{stein} investigated that charged matter 
collapse may produce naked singularity instead of a black hole. Germani and Tsagas \cite{germani} discussed the collapse of magnetized dust in 
Tolman–Bondi model. Recently, Herrera and his collaborators \cite{herrera,prisco} have 
discussed the role of electromagnetic field on the structure scalars and dynamics of self-gravitating objects. Sharif and his 
collaborators \cite{sharif1,sharif2,sharif3} have extended this work for cylindrical and 
plane symmetries.\\

Among the present emphasis in gravitational physics, a theoretical search for an alternative and may be more fundamental theories of gravity 
is of a central attraction. The most 
usual way to modify Einstein’s theory of gravity in a four dimensional context is to add higher curvature terms which 
should arise naturally from the requirement of general coordinate invariance. Such corrections to 
Einstein’s gravity have their natural origin in a fundamental theory like “String Theory”. 
In this context F(R) \cite{paul1,s1,s2,24,25,26}, Gauss-Bonnet (GB) \cite{s1,s2,27,28,29,maeda,mann} or more generally 
Lanczos-Lovelock \cite{lanczos,lovelock} gravity are some of the candidates in higher curvature gravitational theory. 
Higher curvature terms become extremely relevant at
the regime of large curvature. The spacetime curvature inside a collapsing star 
gradually increases as the collapse continues and becomes very large 
near the final state of the collapse. Thus for a collapsing geometry where the curvature becomes very large near the 
final state of the collapse, the higher curvature terms are expected to play a crucial role. Motivated by this idea, the 
collapsing scenarios in the presence of higher curvature gravity have 
been discussed in \cite{paul2,rituparno,nb1,nb3}.\\

In the present work, we investigate the possible effects of electromagnetic field in a 
scalar field collapse in the presence of higher curvature like Gauss-Bonnet gravity. The advantage of 
Gauss-Bonnet (GB) gravity is that the equations of motion do not contain any higher derivative 
terms (higher than two) of the metric and thus leads to ghost free solution. The particular questions that we addressed in this paper 
are the following,

\begin{enumerate}
 \item What are the possible effects of electromagnetic field on scalar field collapse in the presence of Gauss-Bonnet gravity?
 
 \item What is the end product of the collapse, a black hole or a naked singularity?
\end{enumerate}

In order to address the above questions, we consider a ``Scalar-Maxwell-Einstein-Gauss-Bonnet'' theory in four dimension 
where the scalar field is coupled non-minimally to the GB term. It may be mentioned that without the 'non-minimal' 
coupling, the GB contribution in the action does not contribute nontrivially to the field equations in less than 5 dimensions. 
The presence of electromagnetic field forces us to consider 
an anisotropic spacetime model which is discussed in section 2. In section 3, we obtain the exact solution for the metric. 
Section 4 and 5 address the visibility of the singularity produced as a result of the collapse and a matching of the solution with 
an exterior spacetime respectively. We end the paper with some concluding remarks in section 6.\\

\section{The model}
To explore the effect of electromagnetic field on scalar field collapse in presence of higher curvature like Gauss-Bonnet (GB) gravity, 
we consider a ``Scalar-Maxwell-Einstein-Gauss-Bonnet'' theory in four dimensions where the GB term is coupled with the scalar field. This 
coupling guarantees the non topological character of the GB term. The action for this model is given by,
\begin{eqnarray}
 S = \int d^4x \sqrt{-g} \bigg[\frac{R}{2\kappa^2} - \frac{1}{2}g^{\mu\nu}\partial_{\mu}\Phi\partial_{\nu}\Phi - V(\Phi) + \frac{1}{8}\xi(\Phi)G 
 - \frac{1}{4}F_{\mu\nu}F^{\mu\nu}\bigg]
 \label{action}
\end{eqnarray}

where $g$ is the determinant of the metric, $R$ is the Ricci scalar, $1/(2\kappa^2)=M_{p}^2$ is the four dimensional squared Planck mass, 
$G=R^2-4R_{\mu\nu}R^{\mu\nu}+R_{\mu\nu\alpha\beta}
R^{\mu\nu\alpha\beta}$ is the GB term, $\Phi$ denotes the scalar field also endowed with a potential $V(\Phi)$. 
The coupling between scalar field and GB term is symbolized by $\xi(\Phi)$ in the action. The last term in the action denotes 
the electromagnetic field lagrangian where $F_{\mu\nu}$ is the electromagnetic field tensor and is defined by : 
$F_{\mu\nu} = \partial_{\mu}A_{\nu} - \partial_{\nu}A_{\mu}$, $A_{\mu}$ is the electromagnetic four-potential.\\

To obtain the gravitational field equation, we need to determine the energy-momentum tensor for the scalar field ($\Phi$) 
and for the electromagnetic field ($A_{\mu}$) respectively. These stress tensors have the following expressions :
\begin{eqnarray}
 T_{\mu\nu}(\Phi)&=&\frac{2}{\sqrt{-g}} \frac{\delta}{\delta g^{\mu\nu}}\bigg[\sqrt{-g}\bigg(\frac{1}{2}g^{\alpha\beta}\partial_{\alpha}\Phi
 \partial_{\beta}\Phi + V(\Phi)\bigg)\bigg]\nonumber\\
 &=&\bigg[\partial_{\mu}\Phi\partial_{\nu}\Phi - \frac{1}{2}g_{\mu\nu}\partial_{\alpha}\Phi\partial^{\alpha}\Phi - g_{\mu\nu}V(\Phi)\bigg]
 \label{em tensor 1}
 \end{eqnarray}
 
 for the scalar field $\Phi$ and
 
\begin{eqnarray}
 T_{\mu\nu}(A)&=&\frac{2}{\sqrt{-g}} \frac{\delta}{\delta g^{\mu\nu}}\bigg[\frac{1}{4}\sqrt{-g}F_{\alpha\beta}F^{\alpha\beta}\bigg]\nonumber\\
 &=&\bigg[g^{\alpha\beta}F_{\mu\alpha}F_{\nu\beta} - \frac{1}{4}g_{\mu\nu}F_{\alpha\beta}F^{\alpha\beta}\bigg]
 \label{em tensor 2}
\end{eqnarray}

for the electromagnetic field. These above expressions of energy-momentum tensor along with the variation of the action 
 with respect to $g^{\mu\nu}$ leads to the gravitational field equation as follows :
 \begin{eqnarray}
  &\frac{1}{\kappa^2}&G_{\mu\nu} + R_{\mu\nu\alpha\beta}\nabla^{\alpha}\nabla^{\beta}\xi - R_{\mu\nu}\Box\xi 
  + R_{\alpha\nu}\nabla_{\mu}\nabla^{\alpha}\xi + R_{\mu\beta}\nabla^{\beta}\nabla_{\nu}\xi - \frac{1}{2}\nabla_{\mu}\nabla_{\nu}\xi\nonumber\\ 
  &-&\frac{1}{2}g_{\mu\nu}\bigg(2R_{\alpha\beta}\nabla^{\alpha}\nabla^{\beta}\xi - R\Box\xi\bigg) 
  = \bigg(\partial_{\mu}\Phi\partial_{\nu}\Phi - \frac{1}{2}g_{\mu\nu}\partial_{\alpha}\Phi\partial^{\alpha}\Phi - g_{\mu\nu}V(\Phi)\bigg)\nonumber\\ 
  &+&\bigg(g^{\alpha\beta}F_{\mu\alpha}F_{\nu\beta} - \frac{1}{4}g_{\mu\nu}F_{\alpha\beta}F^{\alpha\beta}\bigg)~~~~,
  \label{gravitational equation}
 \end{eqnarray}
 
 where $G_{\mu\nu}$ is the Einstein tensor and $\Box$ ($=g^{\mu\nu}\nabla_{\mu}\nabla_{\nu}$) symbolizes the d'Alembertian 
 operator. It may be noticed that the above equation of motion does not contain any 
 derivative of the metric components higher than two. Similarly the scalar field and the electromagnetic field equations are given by :
 \begin{eqnarray}
  \Box\Phi - V'(\Phi) + \frac{1}{8}\xi'(\Phi)G - \frac{1}{2}F_{\mu\nu}F^{\mu\nu} = 0
  \label{scalar equation}
 \end{eqnarray}
 
 and
 
 \begin{eqnarray}
  \nabla_{\mu}\bigg(\partial^{\mu}A^{\nu} - \partial^{\nu}A^{\mu}\bigg) = 0~~~,
  \label{electromagnetic equation}
 \end{eqnarray}

 where a prime denotes the derivative with respect to the scalar field $\Phi$. It is well known that Einstein-Gauss-Bonnet gravity 
 in four dimensions reduces to standard Einstein gravity, the additional terms actually cancel each other. In the present case, 
 the non-minimal coupling with the scalar field assists the contribution from the GB term survive. It is easy to see, in all the field 
 equations above, that a constant $\xi$ (essentially no coupling) would immediately make the GB contribution trivial.\\
 
 The aim here is to construct a model for a continual collapse. A non static metric ansatz for the interior is taken 
 that fits our purpose. We also consider that the scalar field as well as the electromagnetic field (or gauge field) are homogeneous in space. 
 Under such condition, it can be shown that an anisotropy is essential in order to sustain an electromagnetic field (see Appendix - I). 
 As a candidate of anisotropic model, here we consider the following Bianchi-I metric for interior spacetime,
 \begin{eqnarray}
  ds^2 = -dt^2 + e^{[2\alpha(t) + 2\sigma(t)]} \bigg(dr^2 + r^2d\theta^2\bigg) + e^{[2\alpha(t) - 4\sigma(t)]}dz^2~~~~~.
  \label{ansatz2}
 \end{eqnarray}
 
 It is evident that $e^{\alpha+\sigma}$ and $e^{\alpha-2\sigma}$ 
 are the scale factor along radial direction and along $z$ direction respectively. Hence the Hubble parameter 
 along radial ($H_r$) and $z$ ($H_z$) direction are defined as follows :
 \begin{eqnarray}
  H_r&=&\dot{\alpha} + \dot{\sigma}~,\nonumber\\
  H_z&=&\dot{\alpha} - 2\dot{\sigma}~~~~~~.
  \label{hubble}
 \end{eqnarray}

 Therefore due to the introduction of the gauge field $A_{\mu}(t)$, the spatial isotropy is broken and the 
 deviation from isotropy is controlled by $\sigma(t)$. 
 However the metric in eqn.(\ref{ansatz2}) clearly indicates that $\frac{\partial}{\partial\theta}$, $\frac{\partial}{\partial z}$ 
 are the two killing vector fields for the interior spacetime. Therefore the interior geometry 
 possesses a cylindrical symmetry with $z$ as the longitudinal direction which 
 implies that the anisotropy is generated along the $z$ direction. Hence the component of $A_{\mu}(t)$ can be taken as :
 \begin{eqnarray}
  A_{\mu}(t) = \bigg(0, 0, 0, v(t)\bigg)~~~~~~~~~.
  \nonumber
 \end{eqnarray}
 
 With these above components of $A_{\mu}(t)$, eqn.(\ref{electromagnetic equation}) turns out to be :
 \begin{eqnarray}
  \frac{d}{dt}\bigg[e^{\alpha + 4\sigma} \dot{v}\bigg] = 0~~~~~~~,
  \label{electromagnetic equation anisotropy1}
 \end{eqnarray}
 
 which can be solved to yield
 
 \begin{eqnarray}
  \dot{v}(t) = C e^{[-\alpha(t) - 4\sigma(t)]}~~~~~~~~,
  \label{electromagnetic equation anisotropy2}
 \end{eqnarray}
 
 where $C$ is the constant of integration and an overdot represents the derivative with respect to time ($t$). Using the 
 metric presented in eqn.(\ref{ansatz2}) along with the above solution of $\dot{v}(t)$, eqn.(\ref{gravitational equation} 
 can be simplified and takes the following form :
 \begin{eqnarray}
  \dot{\alpha}^2 = \dot{\sigma}^2 + \frac{\kappa^2}{3}\bigg[V(\Phi) + \frac{\dot{\Phi}^2}{2} + \frac{C^2}{2}e^{-4\alpha-4\sigma}\bigg] 
  - \kappa^2\dot{\xi}\big(\dot{\alpha} - 2\dot{\sigma}\big)\big(\dot{\alpha} + \dot{\sigma}\big)^2~~~~,
  \label{grav eqn anisotropy1}
 \end{eqnarray}

 \begin{eqnarray}
  \ddot{\sigma}&=&-3\dot{\alpha}\dot{\sigma} + \frac{\kappa^2}{3}C^2e^{-4\alpha-4\sigma} - \kappa^2\ddot{\xi} \big(\dot{\alpha}\dot{\sigma} 
  + \dot{\sigma}^2\big)\nonumber\\
  &-&\kappa^2\dot{\xi}\bigg[\dot{\alpha} \big(3\dot{\sigma}^2+\ddot{\alpha}\big) + \dot{\sigma}\big(\ddot{\alpha}+2\ddot{\sigma}\big) 
  + 3\dot{\alpha}^2\dot{\sigma}\bigg]~~~~~~~~,
  \label{grav eqn anisotropy2}
 \end{eqnarray}
 
 \begin{eqnarray}
  \ddot{\alpha}= -3\dot{\alpha}^2 + \kappa^2\bigg[V(\Phi) + \frac{C^2}{6}e^{-4\alpha-4\sigma}\bigg] + \frac{\kappa^2}{2}\ddot{\xi} 
  \big(-\dot{\alpha}^2 + \dot{\sigma}^2\big)\nonumber\\
  + \frac{\kappa^2}{2}\dot{\xi}\bigg[-5\dot{\alpha}^3 + \dot{\alpha}\big(9\dot{\sigma}^2-2\ddot{\alpha}\big) + 4\dot{\sigma}^3 
  + 2\dot{\sigma}\ddot{\sigma}\bigg]~~~~~~~~~.
  \label{grav eqn anisotropy3}
 \end{eqnarray}
 
 Similarly the scalar field equation of motion (see eqn.(\ref{scalar equation})) leads to the following form 
 (recall that the scalar field depends only on the coordinate $t$),

\begin{eqnarray}
 \ddot{\Phi}&=&-3\dot{\alpha}\dot{\Phi} - V'(\Phi) + 3\xi'(\Phi)\big(\dot{\alpha}+\dot{\sigma}\big)\nonumber\\
 &\bigg[&\dot{\alpha}^3 
 - \dot{\alpha}^2\dot{\sigma} + \dot{\alpha}\big(-2\dot{\sigma}^2+\ddot{\alpha}\big) - \dot{\sigma}\big(\ddot{\alpha}+2\ddot{\sigma}\big)\bigg]~~~.
\label{scalar eqn anisotropy} 
\end{eqnarray}

It is evident that due to presence of Gauss-Bonnet term, cubic as well as quartic powers of $\dot{\alpha}$ and 
$\dot{\sigma}$ appear in the above equations. This indicates the non triviality of the Gauss-Bonnet term in presence 
of the coupling function $\xi(\Phi)$ even in four dimension.
\section{Exact solutions : anisotropic collapsing model}
In this section, we present an analytic solution of the field equations 
(eqn.(\ref{grav eqn anisotropy1}) to eqn.(\ref{scalar eqn anisotropy})) and in order to do this, 
we consider a string inspired model \cite{nojiri_55} as follows,
\begin{eqnarray}
 V(\Phi) = V_0 e^{-2\Phi/\Phi_0}~~~~~,
 \nonumber
\end{eqnarray}
and
\begin{eqnarray}
 \xi(\Phi) = \xi_0 e^{2\Phi/\Phi_0}~~~~~~,
\end{eqnarray}

where $V_0$, $\xi_0$ and $\Phi_0$ are the parameters of the model. 
With these forms of $V(\Phi)$ and $\xi(\Phi)$, eqn.(\ref{grav eqn anisotropy1}) to eqn.(\ref{scalar eqn anisotropy}) turn out be
\begin{eqnarray}
 \dot{\alpha}^2&=&\dot{\sigma}^2 + \frac{\kappa^2}{3}\bigg[V_0 e^{-2\Phi/\Phi_0} + \frac{\dot{\Phi}^2}{2} 
 + \frac{C^2}{2}e^{-4\alpha-4\sigma}\bigg]\nonumber\\ 
  &-&\frac{2\kappa^2\xi_0}{\Phi_0}e^{2\Phi/\Phi_0}\dot{\Phi}\big(\dot{\alpha} - 2\dot{\sigma}\big)\big(\dot{\alpha} + \dot{\sigma}\big)^2~~~~,
 \label{new1}
\end{eqnarray}

\begin{eqnarray}
 \ddot{\sigma}&=&-3\dot{\alpha}\dot{\sigma} + \frac{\kappa^2}{3}C^2e^{-4\alpha-4\sigma} - \kappa^2\xi_0e^{2\Phi/\Phi_0} 
 \big(\frac{2}{\Phi_0}\ddot{\Phi}+\frac{4}{\Phi_0^2}\dot{\Phi}^2\big)\big(\dot{\alpha}\dot{\sigma} 
  + \dot{\sigma}^2\big)\nonumber\\
  &-&\frac{2\kappa^2\xi_0}{\Phi_0}e^{2\Phi/\Phi_0}\dot{\Phi}\bigg[\dot{\alpha} \big(3\dot{\sigma}^2+\ddot{\alpha}\big) 
  + \dot{\sigma}\big(\ddot{\alpha}+2\ddot{\sigma}\big) 
  + 3\dot{\alpha}^2\dot{\sigma}\bigg]~~~~~~~~,
 \label{new2}
\end{eqnarray}

\begin{eqnarray}
 \ddot{\alpha}&=&-3\dot{\alpha}^2 + \kappa^2\bigg[V_0e^{-2\Phi/\Phi_0} + \frac{C^2}{6}e^{-4\alpha-4\sigma}\bigg] 
 + \kappa^2\xi_0e^{2\Phi/\Phi_0}\big(\frac{\ddot{\Phi}}{\Phi_0}+\frac{2\dot{\Phi}^2}{\Phi_0^2}\big) \nonumber\\
  &\big(&-\dot{\alpha}^2 + \dot{\sigma}^2\big)
  + \frac{\kappa^2\xi_0}{\Phi_0}e^{2\Phi/\Phi_0}\dot{\Phi}\bigg[-5\dot{\alpha}^3 
  + \dot{\alpha}\big(9\dot{\sigma}^2-2\ddot{\alpha}\big) + 4\dot{\sigma}^3 
  + 2\dot{\sigma}\ddot{\sigma}\bigg]
 \label{new3}
\end{eqnarray}
and 
\begin{eqnarray}
 \ddot{\Phi}&=&-3\dot{\alpha}\dot{\Phi} + \frac{2V_0}{\Phi_0}e^{-2\Phi/\Phi_0} 
  + \frac{6\xi_0}{\Phi_0}e^{2\Phi/\Phi_0}\big(\dot{\alpha}+\dot{\sigma}\big)\nonumber\\ 
 &\bigg[&\dot{\alpha}^3 
 - \dot{\alpha}^2\dot{\sigma} + \dot{\alpha}\big(-2\dot{\sigma}^2+\ddot{\alpha}\big) - \dot{\sigma}\big(\ddot{\alpha}+2\ddot{\sigma}\big)\bigg]
 \label{new4}
\end{eqnarray}

respectively. Here we are interested on the collapsing solutions where the volume 
of the two cylinder (recall that the interior spacetime has a cylindrical symmetry) 
decreases monotonically with time. Keeping this in mind, the above four equations (eqn.(\ref{new1}), eqn.(\ref{new2}), eqn.(\ref{new3}), 
eqn.(\ref{new4})) are solved for $\alpha(t)$, $\sigma(t)$, $\Phi(t)$ and the solutions are the following:
\begin{equation}
 e^{\alpha(t)} \propto (t_0-t)^{\alpha_0},
 \label{sol_alpha}
\end{equation}

\begin{eqnarray}
 e^{\sigma(t)} \propto (t_0-t)^{\sigma_0},
 \label{sol_sigma}
\end{eqnarray}

and

\begin{equation}
 \Phi(t) = \Phi_0 \ln {\bigg[\frac{1}{\kappa}\big(t_0 - t\big)\bigg]}
 \label{sol_scalar}
\end{equation}

where $t_0$ is a constant of integration. The constants $\alpha_0$, $\sigma_0$ and $C$ $\big($appeared in the solution of electromagnetic field, 
see eqn.(\ref{electromagnetic equation anisotropy2})$\big)$ are related to $V_0$, $\xi_0$ (taken as greater than zero, which is consistent 
with the local astronomical tests \cite{astronomy}) and $\Phi_0$ through the following four relations, 
\begin{eqnarray}
\alpha_0 + \sigma_0 = \frac{1}{2},
\label{condition_1}
\end{eqnarray}
\begin{eqnarray}
 \alpha_0^2 = \sigma_0^2 + \frac{\kappa^2}{3}\big(V_0\kappa^2+\frac{1}{2}\Phi_0^2\big) + \frac{\kappa^2}{6}C^2 
 + \frac{\xi_0}{2}\big(2\sigma_0-\alpha_0\big),
 \label{condition_2}
 \end{eqnarray}
 \begin{eqnarray}
 \frac{\kappa^2}{3}C^2 = \sigma_0\big(3\alpha_0-1\big)\big(1+\xi_0\big),
 \label{condition_3}
 \end{eqnarray}
 and
 \begin{eqnarray}
 \alpha_0&=&3\alpha_0^2 - \kappa^4V_0 - \frac{\kappa^2}{6}C^2\nonumber\\
 &+&\xi_0\big(5\alpha_0^3-9\alpha_0\sigma_0^2-3\alpha_0^2-4\sigma_0^3-3\sigma_0^2\big).
 \label{condition_4}
\end{eqnarray}

Eqn.(\ref{sol_sigma}) depicts that the exponent of $e^{\sigma(t)}$ (effectively $\sigma_0$) determines the strength of anisotropy of the 
spacetime. Further it may be observed from eqn.(\ref{condition_3}) that for $C\neq0$, the anisotropy factor $\sigma_0$ cannot be zero. 
These reflect the fact that the presence of the time dependent electromagnetic field calls for an anisotropy in the spacetime geometry. 
However for $C=0$ (i.e in the absence of the electromagnetic field), 
the spacetime either becomes isotropic ($\sigma_0=0$) or possesses a certain anisotropy with $\sigma_0=1/6$. Later we discuss 
the possible consequences of such situations on the collapsing phenomena. 
The solutions of $\alpha(t)$, $\sigma(t)$ (in eqn.(\ref{sol_alpha}), eqn.(\ref{sol_sigma})) immediately lead 
to the evolution of scale factor along radial and longitudinal directions as,
\begin{eqnarray}
 a_r(t)&=&e^{[\alpha(t)+\sigma(t)]}\nonumber\\
 &=&B^{(r)}_0(t_0 - t)
 \label{sol_scale1}
 \end{eqnarray}
 and
 \begin{eqnarray}
 a_z(t)&=&e^{[\alpha(t)-2\sigma(t)]}\nonumber\\
 &=&B^{(z)}_0(t_0 - t)^{\frac{1}{2}-3\sigma_0}
 \label{sol_scale2}
\end{eqnarray}

respectively where $B^{(r)}_0$ and $B^{(z)}_0$ are integration constants. To derive the above two expressions, we use 
eqn.(\ref{condition_1}). The expression of $a_r(t)$ (see eqn.(\ref{sol_scale1})) 
clearly reveals that $ra_r(t)$ decreases monotonically with time. Therefore, the volume 
of the cylinder of the scalar field collapses with time and goes to zero at $t\rightarrow t_0$, giving rise 
to a finite time zero proper volume singularity. On the other hand, the evolution of the scale factor along longitudinal direction $a_z(t)$ 
depends on the anisotropy factor $\sigma_0$. For $\sigma_0 < 1/6$, $a_z(t)$ decreases monotonically with time and 
goes to zero at $t\rightarrow t_0$, while the condition $\sigma_0 > 1/6$ entails that $a_z(t)$ continually increases and as a result, diverges 
at $t\rightarrow t_0$. Therefore the singularity appeared at $t\rightarrow t_0$ is a point singularity 
for $\sigma_0 < 1/6$ while for the other condition, the collapse ends to a line singularity. This directs us to argue that 
the nature of the singularity depends entirely on the strength of anisotropy of the spacetime with the limiting situation as 
defined by $\sigma_0=1/6$. For such limiting case, $a_z(t)$ becomes constant (finite) which in turn leads the collapse 
to a ``finite line singularity''. Further recall from eqn.(\ref{condition_3}) that this 
limiting condition corresponds to $C=0$. Therefore the final fate of the collapsing scalar field 
in absence of the electromagnetic field is depicted by such ``finite line singularity''.\\

In order to investigate whether the singularity is a curvature singularity or just an artifact of coordinate choice, one must 
look into the behaviour of Kretschmann curvature scalar ($K = R_{\mu\nu\alpha\beta}R^{\mu\nu\alpha\beta}$) at $t\rightarrow t_0$. For the metric 
presented in eqn.(\ref{ansatz2}), $K$ has the following expression,\\
\begin{equation}
 K = 4\bigg(\frac{\dot{a}_r}{a_r}\bigg)^4 + 8\bigg(\frac{\dot{a}_r\dot{a}_z}{a_ra_z}\bigg)^2 + 4\bigg[2\bigg(\frac{\ddot{a}_r}{a_r}\bigg) 
 + \bigg(\frac{\ddot{a}_z}{a_z}\bigg)\bigg]^2
 \label{curvature_scalar1}
\end{equation}

Using the solutions of $a_r(t)$ and $a_z(t)$ (see eqn. (\ref{sol_scale1}) and eqn.(\ref{sol_scale2})), the 
above expression of $K$ can be simplified as,\\
\begin{eqnarray}
 K = \frac{4}{\big(t_0 - t\big)^4}\bigg[\frac{1}{16} + \frac{1}{2}(\alpha_0-2\sigma_0)^2 + \bigg((\alpha_0-2\sigma_0)^2 - (\alpha_0-2\sigma_0) 
 -\frac{1}{2}\bigg)^2\bigg]
 \label{curvature_scalar2}
\end{eqnarray}

It is clear from eqn.(\ref{curvature_scalar2}) that the Kretschmann scalar diverges at $t\rightarrow t_0$ and 
thus the collapsing cylinder discussed here ends up in a curvature singularity.\\

\begin{figure}[!h]
\begin{center}
 \centering
 \includegraphics[width=3.0in,height=2.0in]{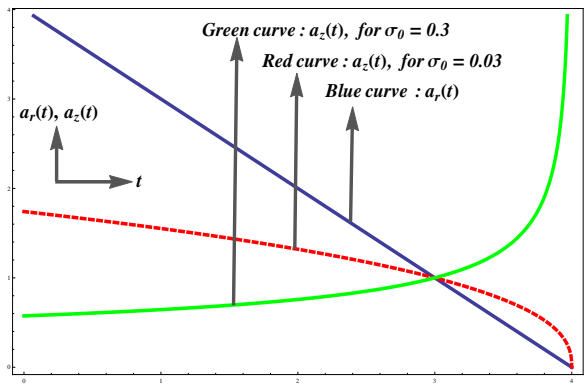}
 \caption{$a_r(t)$, $a_z(t)$ versus $t$}
 \label{plot scale}
\end{center}
\end{figure}

From eqn.(\ref{sol_scale1}) and eqn.(\ref{sol_scale2}), we obtain the plot (figure (\ref{plot scale})) of $a_r(t)$, $a_z(t)$ versus $t$. 

Figure (\ref{plot scale}) clearly demonstrates that $a_r(t)$ decreases with time linearly and goes to zero as $t$ tends to $t_0$. 
On the other hand $a_z(t)$ decreases almost uniformly until $t$ approaches a value close to $t_0$, where it hurries towards a zero proper volume singularity. 
This collapsing behaviour of $a_z(t)$ is shown in the dashed curve where $\sigma_0$ is taken as $0.03$. On the other hand, 
the green solid curve depicts the diverging character of $a_z(t)$ for $\sigma_0 = 0.3$.\\

\section{Visibility of the singularity}
The visibility of curvature singularity to an exterior observer depends on the formation 
of an apparent horizon. The condition for such a surface is given by
\begin{eqnarray}
 g^{\mu\nu} Z_{,\mu} Z_{,\nu}\bigg|_{r_{ah},t_{ah}} = 0
 \label{app hor 1}
\end{eqnarray}

where $Z$ is the proper radius of the two cylinder, given by $ra_r(t)$ in the present case, $r_{ah}$ and 
$t_{ah}$ being the comoving radial coordinate and time of formation of the apparent horizon respectively. 
Using the form of $g^{\mu\nu}$ presented in eqn.(\ref{ansatz2}), above expression can be simplified 
and turns out to be,
\begin{eqnarray}
 r_{ah}^2 \dot{a}_r(t_{ah})^2 = 1
 \label{app hor 2}
\end{eqnarray}

where we use $a_r(t)=e^{[\alpha(t)+\sigma(t)]}$. Due to the solution of $a_r(t)$, eqn.(\ref{app hor 2}) takes the following form :
\begin{eqnarray}
 \big[t_0 - t_{ah}\big] = \frac{1}{4}B^{(r)}_0 r_{ah}^2~~~~~~~~.
 \label{app hor 3}
\end{eqnarray}
The above expression clearly demonstrates that $t_{ah}$ is less than $t_0$ (i.e. $t_{ah} < t_0$). 
Therefore the formation of apparent horizon lags behind than the formation of singularity. Thus, the curvature singularity 
discussed here is always covered from an exterior observer by the apparent horizon. 
At this stage, it may be mentioned 
that the singularity formed is not a central singularity, it is formed at any value of $r$ within 
the distribution. Such a singularity in general relativity is always covered by a horizon \cite{31_NS}.\\

\section{Matching of the interior spacetime with an exterior geometry}
To complete the model, the interior spacetime geometry of the collapsing scalar field cylindrical cloud 
(recall that the interior geometry is cylindrically symmetric) needs to be matched to 
an exterior geometry. For the required matching, the Israel conditions are used, 
where the metric coefficients and extrinsic curvatures (first and second fundamental forms 
respectively) are matched at the boundary of the cylinder \cite{nolan}. At this stage, it deserves mention that 
the Gauss-Bonnet term (controlled by the coupling function $\xi(\Phi)$) generates an effective energy momentum tensor which 
can not be zero (since it arises effectively
from spacetime curvature) at the exterior and hence the presence of Gauss-Bonnet gravity spoils the matching of the collapsing 
interior spacetime with a vacuum exterior geometry. Such spoiling of matching (with a vacuum exterior) 
due to the presence of Gauss-Bonnet gravity can also be found in the previous literature \cite{paul2}. In addition, the energy 
momentum tensor carried by the electromagnetic field will further lead to an inconsistency if the interior collapsing cloud is matched 
with a vacuum exterior. For instance, since vacuum has zero electromagnetic (em) field, such a matching would lead to a discontinuity 
in the em field, which means a delta function in
the gradient of the em field. As a consequence, there will appear square of a delta function in the
stress-energy, which is definitely an inconsistency. Keeping these in mind, here we match 
the interior geometry with a generalized cylindrically symmetric exterior spacetime \cite{nolan,kompaneets,jordan} at 
the boundary hypersurface $\Sigma$ 
given by $r = r_0$. The metric inside and outside of $\Sigma$ are given by,
\begin{eqnarray}
 ds_{-}^2 = -dt^2 + e^{[2\alpha(t) + 2\sigma(t)]} \bigg(dr^2 + r^2d\theta^2\bigg) + e^{[2\alpha(t) - 4\sigma(t)]}dz^2
 \label{inside metric}
\end{eqnarray}
and
\begin{eqnarray}
 ds_{+}^2 = e^{2(\Upsilon-\Psi)}\big(-dT^2+d\rho^2\big) + R^2e^{-2\Psi}d\theta^2 + e^{2\Psi}\big(dz+Wd\theta\big)^2
 \label{outside metric}
\end{eqnarray}
respectively, where $T,\rho,\theta$ and $z$ are the exterior coordinates and $\Upsilon$, $\Psi$, $R$, $W$ are 
functions of $T$ and $\rho$. Therefore $\frac{\partial}{\partial\theta}$ and $\frac{\partial}{\partial z}$ 
are the killing vector fields of the exterior spacetime which yields a cylindrical 
symmetry in the exterior. The same hypersurface $\Sigma$ can alternatively be defined by the exterior 
coordinates as $T = T(t)$ and $\rho = \rho(t)$. Then the metrics on $\Sigma$ from inside and 
outside coordinates turn out to be,
\begin{eqnarray}
 ds_{-,\Sigma}^2 = -dt^2 + e^{[2\alpha(t) + 2\sigma(t)]} r_0^2d\theta^2 + e^{[2\alpha(t) - 4\sigma(t)]}dz^2
 \nonumber
\end{eqnarray}
and
\begin{eqnarray}
 ds_{+,\Sigma}^2 = e^{2(\Upsilon_{\Sigma} - \Psi_{\Sigma})}\big(-\dot{T}^2 + \dot{\rho}^2\big) dt^2 
 + R_{\Sigma}^2e^{-2\Psi_{\Sigma}}d\theta^2 + e^{2\Psi_{\Sigma}}\big(dz + W_{\Sigma}d\theta\big)^2
 \nonumber
\end{eqnarray}
where $\Upsilon_{\Sigma}(t)$ $\big(=\Upsilon(T(t),\rho(t))\big)$, $\Psi_{\Sigma}(t)$, $R_{\Sigma}(t)$ and $W_{\Sigma}(t)$ 
are the respective functions defined on $\Sigma$ and dot represents $\frac{d}{dt}$. Matching the first fundamental 
form on $\Sigma$ (i.e. $ds_{-,\Sigma}^2 = ds_{+,\Sigma}^2$) yields the following conditions :
\begin{eqnarray}
 e^{2(\Upsilon_{\Sigma} - \Psi_{\Sigma})}\big(\dot{T}^2 - \dot{\rho}^2\big) = 1~~~,
 \label{con 1}
\end{eqnarray}

\begin{eqnarray}
 e^{\Psi_{\Sigma}(t)}&=&a_z(t)\nonumber\\
 &=&B^{(z)}_0(t_0 - t)^{\frac{1}{2}-3\sigma_0}~~~,
 \label{con 2}
\end{eqnarray}

\begin{eqnarray}
 R_{\Sigma}(t)&=&r_0 a_r(t)a_z(t)\nonumber\\
 &=&r_0 B^{(r)}_0 B^{(z)}_0(t_0 - t)^{\frac{3}{2}-3\sigma_0}
 \label{con 3}
\end{eqnarray}
and
\begin{eqnarray}
 W_{\Sigma}(t) = 0.
 \label{con 4}
\end{eqnarray}

In order to match the second fundamental form, we calculate the normal of the hypersurface $\Sigma$ 
from inside ($\vec{n}^{-} = n^{-}_t$, $n^{-}_r$, $n^{-}_{\theta}$, $n^{-}_z$) and outside 
($\vec{n}^{+} = n^{+}_T$, $n^{+}_{\rho}$, $n^{+}_{\theta}$, $n^{+}_z$) coordinates as follows :
\begin{eqnarray}
 n^{-}_t = 0~~,~~~~~~~~n^{-}_r =  a(t)~~,~~~~~~~n^{-}_{\theta} = n^{-}_z = 0\label{inside normal}
\end{eqnarray}
and
\begin{eqnarray}
 n^{+}_T = \frac{e^{(\Upsilon_{\Sigma} - \Psi_{\Sigma})}\dot{\rho}}{\sqrt{\dot{T}^2-\dot{\rho}^2}}~~,\nonumber
 \end{eqnarray}
 \begin{eqnarray}
 n^{+}_{\rho} = \frac{e^{(\Upsilon_{\Sigma} - \Psi_{\Sigma})}\dot{T}}{\sqrt{\dot{T}^2-\dot{\rho}^2}}~~,\nonumber
 \end{eqnarray}
 \begin{eqnarray}
 n^{+}_{\theta} = n^{+}_z = 0.
\label{outside normal}
\end{eqnarray}

The above expressions of $\vec{n}^{-}$ and $\vec{n}^{+}$ leads to the extrinsic 
curvature of $\Sigma$ from interior and exterior coordinates respectively, and are given by,
\begin{eqnarray}
 K_{tt}^- = 0~~,~~~~~~~~~K_{\theta\theta}^- = r_0a_r(t)~~,~~~~~~~K_{zz}^- = 0
 \label{inside extrinsic}
\end{eqnarray}
 (all the other components of $K_{\mu\nu}^{-}$ are zero) from interior metric, and
\begin{eqnarray}
 K_{tt}^+ = e^{(\Upsilon_{\Sigma}-\Psi_{\Sigma})} \sqrt{\dot{T}^2 - \dot{\rho}^2} 
 \bigg[\big(\Psi_{\rho}\dot{T}-\Psi_T\dot{\rho}\big) - \big(\Upsilon_{\rho}\dot{T}-\Upsilon_T\dot{\rho}\big)\bigg]~~,
 \nonumber
\end{eqnarray}
\begin{eqnarray}
 K_{\theta\theta}^+ =  \frac{R_{\Sigma}e^{-(\Upsilon_{\Sigma}+\Psi_{\Sigma})}}{\sqrt{\dot{T}^2 - \dot{\rho}^2}} 
 \bigg[\big(R_{\rho}\dot{T}-R_T\dot{\rho}\big) - R_{\Sigma}\big(\Psi_{\rho}\dot{T}-\Psi_T\dot{\rho}\big)\bigg]~~~~,
 \nonumber
\end{eqnarray}
\begin{eqnarray}
 K_{zz}^+ = \frac{e^{-(\Upsilon_{\Sigma}-3\Psi_{\Sigma})}}{\sqrt{\dot{T}^2 - \dot{\rho}^2}} 
 \bigg[\Psi_{\rho}\dot{T}-\Psi_T\dot{\rho}\bigg]~~~~,
 \nonumber
\end{eqnarray}
\begin{eqnarray}
 K_{z\theta}^+&=&K_{\theta z}^+\nonumber\\
 &=&\frac{e^{-(\Upsilon_{\Sigma}-3\Psi_{\Sigma})}}{\sqrt{\dot{T}^2 - \dot{\rho}^2}} \bigg[W_{\rho}\dot{T}-W_T\dot{\rho}\bigg]  
 \label{outside extrinsic}
\end{eqnarray}

 (all the other components of $K_{\mu\nu}^{+}$ are zero) from exterior metric, where the subscription denotes 
 the respective derivative on the hypersurface $\Sigma$, such as $R_T = \frac{\partial R}{\partial T}\bigg|_{\Sigma}$.\\

The equality of the extrinsic curvatures at $\Sigma$ from both sides is therefore equivalent to the following conditions :\\

\begin{eqnarray}
 \bigg[R_{\rho}\dot{T} - R_T\dot{\rho}\big)\bigg] 
 \frac{R_{\Sigma}e^{-(\Upsilon_{\Sigma}+\Psi_{\Sigma})}}{\sqrt{\dot{T}^2 - \dot{\rho}^2}}&=&r_0 a_r(t)\nonumber\\
 &=&r_0B^{(r)}_0(t_0 - t)~~~~~,
 \label{con 5}
\end{eqnarray}

\begin{eqnarray}
 \bigg[\Upsilon_{\rho}\dot{T} - \Upsilon_T\dot{\rho}\bigg] = 0~~~~,
 \label{con 6}
\end{eqnarray}

\begin{eqnarray}
 \bigg[\Psi_{\rho}\dot{T} - \Psi_T\dot{\rho}\bigg] = 0~~~~,
 \label{con 7}
\end{eqnarray}
and
\begin{eqnarray}
 \bigg[W_{\rho}\dot{T} - W_T\dot{\rho}\bigg] = 0.
 \label{con 8}
\end{eqnarray}

Eqn.(\ref{con 5}) can be further simplified by using the conditions obtained in eqn.(\ref{con 1}), eqn.(\ref{con 2}), eqn.(\ref{con 3}) 
and finally we obtain the following expression
\begin{eqnarray}
 \bigg[R_{\rho}\dot{T} - R_T\dot{\rho}\bigg]&=&\frac{e^{2\Upsilon_{\Sigma}}}{a_z(t)}\nonumber\\
 &=&\frac{e^{2\Upsilon_{\Sigma}}}{B^{(z)}_0(t_0 - t)^{\frac{1}{2}-3\sigma_0}}
 \label{con 9}
\end{eqnarray}

The above four relations along with eqn.(\ref{con 1}) to eqn.(\ref{con 4}) completely 
specify the matching at the boundary of the collapsing scalar field with an exterior cylindrically symmetric geometry.\\

\section{Conclusion}
We consider a ``Scalar-Maxwell-Einstein-Gauss-Bonnet'' theory in four dimensions where the scalar field 
couples non-minimally with the Gauss-Bonnet (GB) term. This coupling with the scalar field guarantees the 
non topological character of the GB term. In this higher curvature theory, we examine the possible effects 
of the electromagnetic field on scalar field collapse.\\

The presence of electromagnetic field requires an anisotropic metric. We consider 
a special Bianchi-I metric $\big($which possesses 
a cylindrical symmetry, the radial scale factor ($a_r(t)$) is different form the longitudinal scale 
factor ($a_z(t)$)$\big)$ as a candidate of an anisotropic model. With the aforementioned metric, an exact 
solution is obtained for the spacetime geometry, which clearly 
reveals that the radius of a two cylinder decreases monotonically with time. Therefore, the volume 
of the cylinder of the scalar field collapses and goes to zero at a finite time ($t_0$) leading 
to a zero proper volume singularity. From the behaviour of Kretschmann scalar, it is found that the singularity formed as a result of the 
collapse is a finite time curvature singularity.\\

On the other hand, the evolution of the longitudinal scale factor indicates that for $\sigma_0 < 1/6$, $a_z(t)$ decreases with time 
and goes to zero at $t \rightarrow t_0$ while the condition $\sigma_0 > 1/6$ makes $a_z(t)$ an 
increasing function of time and as a consequence, diverges at $t \rightarrow t_0$. The parameter $\sigma_0$ is essentially 
determined by the Gauss-Bonnet coupling (with the scalar field) $\xi_0$ and the parameters $V_0$, $\Phi_0$. However such 
collapsing or diverging behaviours of $a_z(t)$ demonstrate that the singularity 
we discussed here is point like or line like depending on the condition whether $\sigma_0 < 1/6$ or $\sigma_0 > 1/6$ respectively. 
Moreover, it may be mentioned that the parameter $\sigma_0$ actually regulates the strength of the spacetime anisotropy. 
Therefore it can be argued 
that in the present context, the pattern of the singularity (point like or line like) is controlled by the strength of anisotropy of the spacetime 
with the limiting situation is defined by $\sigma_0=1/6$. For such limiting case, $a_z(t)$ becomes constant (finite) which in turn leads the collapse 
to a ``finite line singularity''. Further this 
limiting condition corresponds to $C=0$ (see eqn.(\ref{condition_3})). Therefore the final state of the scalar field collapse 
in absence of the electromagnetic field is demonstrated by such ``finite line singularity''.\\

The visibility of curvature singularity to an exterior observer depends on apparent horizon. The formation of apparent horizon is investigated and 
it turns out that the apparent horizon forms before the collapsing cloud hits to singularity. 
Therefore the curvature singularity is hidden from exterior by an apparent horizon. 
Here, it deserves mentioning that the singularity is independent of the radial coordinate $r$ and it is covered by 
a horizon. This result is consistent with the result obtained by Joshi {\it et al} \cite{31_NS} that unless one has a 
central singularity, it can not be a naked singularity. It is interesting to note that the result 
obtained in the present work in the presence of Gauss-Bonnet term is completely consistent with the corresponding GR result. Such 
consistency between Gauss-Bonnet gravity and Einstein's GR is also in agreement with \cite{paul2}.\\

Finally, we match the interior collapsing spacetime geometry with a generalized cylindrically symmetric exterior geometry at 
the boundary of the cloud ($\Sigma$). For this matching, the Israel junction conditions are used where the metric 
coefficients and extrinsic curvatures are matched on $\Sigma$.\\
\newpage

\section*{Appendix - I: Situation of isotropic spacetime}
The non static isotropic metric ansatz is taken as,
 \begin{eqnarray}
  ds^2 = -dt^2 + a^2(t) \bigg[dr^2 + r^2d\theta^2 + dz^2\bigg]
  \label{ansatz1}
 \end{eqnarray}
 with $a(t)$ is the scale factor of the spacetime characterized by the coordinates $t$ ($=x^0$), $r$ ($=x^1$), 
 $\theta$ ($=x^2$) and $z$ ($=x^3$) where $t$ is the timelike one. Moreover the scalar field and the electromagnetic field are considered to be 
 dependent only on $t$. Therefore $F_{\mu\nu}$ has three non zero independent components : $F_{01}$, $F_{02}$ and $F_{03}$. With these non 
 zero components of $F_{\mu\nu}$, we obtain various components of $T_{\mu\nu}(A)$ from eqn.(\ref{em tensor 2}) and are given by,
 \begin{eqnarray}
  T_{00}&=&\frac{1}{2}\bigg[F_{01}F^{01} + F_{02}F^{02} + F_{03}F^{03}\bigg]\nonumber\\
  T_{11}&=&-\frac{1}{2}a^2\bigg[F_{01}F^{01} - F_{02}F^{02} - F_{03}F^{03}\bigg]\nonumber\\
  T_{22}&=&-\frac{1}{2}a^2\bigg[-F_{01}F^{01} + F_{02}F^{02} - F_{03}F^{03}\bigg]\nonumber\\
  T_{33}&=&-\frac{1}{2}a^2\bigg[-F_{01}F^{01} - F_{02}F^{02} + F_{03}F^{03}\bigg]\nonumber\\
  T_{10}&=&T_{20} = T_{30} = 0\nonumber\\  
  T_{12}&=&-a^2 F_{01}F^{02}~~~~~,~~~~T_{13} = -a^2 F_{01}F^{03}~~~~,~~~T_{23} = -a^2 F_{02}F^{03}
  \label{1}
 \end{eqnarray}
 
 Using the above expressions of $T_{\mu\nu}(A)$, the non diagonal components of gravitational equation are simplified to the following 
 form :\\
 
 \begin{eqnarray}
  F_{01}F^{02} = F_{01}F^{03} = F_{02}F^{03} = 0
  \label{2}
 \end{eqnarray}
 
 which has the solution as $F_{01} = F_{02} = F_{03} = 0$. Thus a spatially flat isotropic spacetime cannot support the time dependent 
 electromagnetic field. However a Bianchi-I spacetime, although it is spatially flat, can sustain the gauge field by virtue of its anisotropy.

\end{document}